\documentclass[a4paper]{jpconf}

\usepackage{graphicx}
\usepackage{amsmath}
\usepackage{amssymb}

\begin{document}

\title{On Boiling of Crude Oil under Elevated Pressure}

\author{Anastasiya V Pimenova$^1$ and Denis S Goldobin$^{1,2,3}$}
\address{$^1$Institute of Continuous Media Mechanics, UB RAS,
             Perm 614013, Russia}
\address{$^2$Department of Theoretical Physics, Perm State University,
             Perm 614990, Russia}
\address{$^3$Department of Mathematics, University of Leicester,
             Leicester LE1 7RH, UK}
\ead{Anastasiya.Pimenova@gmail.com, Denis.Goldobin@gmail.com}

\begin{abstract}
We construct a thermodynamic model for theoretical calculation of the boiling process of multicomponent mixtures of hydrocarbons ({\it e.g.}, crude oil). The model governs kinetics of the mixture composition in the course of the distillation process along with the boiling temperature increase. The model heavily relies on the theory of dilute solutions of gases in liquids. Importantly, our results are applicable for modelling the process under elevated pressure (while the empiric models for oil cracking are not scalable to the case of extreme pressure), such as in an oil field heated by lava intrusions.
\end{abstract}

The problem of boiling of crude oil fields by lava intrusions is an important geophysical problem. The impact of the tectonic activity on oil fields not only controls the composition and geological characteristics of the deposits experienced this impact, but may have also an important influence on the climate. For instance, the Great Permian Extinction contemporary to the formation of the Ural mountains is hypothetically associated with massive release of light hydrocarbon gases into the atmosphere. The today's assessments of the amount of methane hydrates on the Earth are essentially more modest than the earlier ones and suggest that the destabilisation of methane hydrates could not be a sufficient source of these gases. Meanwhile, the oil fields boiled by lava intrusions in the region of forming Ural mountains could be this source. Indeed, the oil deposits in this region and the adjoin part of the contemporary basin of the Volga river are features by a nearly absent light fraction.

For modelling this process, one needs a thermodynamic model for crude oil boiling at elevated pressure. The empiric data on the oil cracking are insufficient here, since the technological process of the oil cracking is performed at atmospheric or lower pressure. In what follows we suggest the physical model for boiling of multicomponent mixtures of hydrocarbons, derive  mathematical model governing the kinetics of the boiling process and evaluate its physical parameters. The derived model can be employed for modelling of the boiling process under pressure not significantly exceeding $100\,\mathrm{atm}$, which corresponds to $1\,\mathrm{km}$ depth in sediments.

\section{Boiling of multicomponent mixture of hydrocarbons}
\subsection{Basic physical model}
As a first step, we need to determine thermodynamic equilibrium composition of the vapour phase above the liquid multicomponent mixture of hydrocarbons at its boiling temperature. This can be achieved on the basis of the following theses:
\\
$\bullet$\ The components are indexed by the number $m$ of C-atoms in their molecules, and the difference in thermodynamic characteristics of the interaction of the molecules with the liquid phase between various hydrocarbons with the same number $m$ is neglected. In support to this model reduction, let us notice that the boiling temperatures of aromatic hydrocarbons, cycloalkanes, and isomers of alkanes with $m$ C-atoms are between the boiling points of n-alkanes with $(m-1)$ and $(m+1)$ with rare exceptions for isomers which are untypical for oil up to $m\approx12$.
\\
$\bullet$\ Within the framework of our model reduction, the oil composition is described by the molar fraction $X(m)$ (or $X^{[m]}$ in our notations below) of the molecules with $m$ C-atoms.
\\
$\bullet$\ As the interaction between hydrocarbon molecules is short-range and, therefore, a molecule in liquid oil interacts dominantly with the short nearest segments of nearby passing chains of other molecules, the interaction potential energy will be similar, {\it e.g.}, for segments of the octane's chain and the dodecane's chain. Hence, as a rough approximation, the potential energy of interaction of a molecule with the oil liquid is assumed to be independent of the oil composition.
\\
$\bullet$\ The molar fraction of each component in oil is small; therefore, one can consider the relation between the vapour pressure of this component above the liquid oil and its concentration in oil adopting the theory of solubility of gases in liquids for infinitely dilute solutions---specifically, the scaled particle theory~\cite{Pierotti-1976}.
\\
$\bullet$\ The vapour phase is considered as a perfect gas, which is reasonable as it is mainly composed by light hydrocarbons and the reference pressure is $100\,\mathrm{atm}$, which corresponds to the depth of oil formation about $1\,\mathrm{km}$. (In appendix of~\cite{Goldobin-Brilliantov-2011}, one can see that the van der Waals corrections to the gas solubility are small for pressure about $100\,\mathrm{atm}$.)

According to the scaled particle theory the saturated vapour pressure of specie $m$ dilutely dissolved in the liquid is
\begin{equation}
P_\mathrm{vap}^{[m]}=
\frac{RT}{v_\mathrm{liq}}X^{[m]}
 \exp\left[\frac{G_c^{[m]}+G_i^{[m]}}{k_BT}\right],
\label{eq-MCM-01}
\end{equation}
where $R=8.314\,\mathrm{J/(mol\cdot K)}$ is the universal gas constant, $T$ is temperature, $v_\mathrm{liq}$ is the volume of $1\,\mathrm{mol}$ of the liquid matter, $k_B=1.38\cdot10^{-23}\,\mathrm{J/K}$ is the Boltzmann constant, $G_c$ is the cavity formation work for the guest molecule in the solvent, $G_i$ is the internal energy of interaction of the solute molecule with the ambient solvent molecules. The cavity formation work reads~\cite{Pierotti-1976}
\begin{equation}
\frac{G_c^{[m]}}{k_BT}=-\ln(1-y)+\frac{3y}{1-y}
\frac{\sigma^{[m]}}{\sigma_\mathrm{liq}}
 +\left[\frac{3y}{1-y}+\frac{9}{2}\left(\frac{y}{1-y}\right)^2\right]
 \left(\frac{\sigma^{[m]}}{\sigma_\mathrm{liq}}\right)^2
 +\frac{yP}{n_\mathrm{liq}k_BT}
 \left(\frac{\sigma^{[m]}}{\sigma_\mathrm{liq}}\right)^3,
\label{eq-MCM-02}
\end{equation}
where $\sigma^{[m]}$ is the effective diameter of the solute molecules, $\sigma_\mathrm{liq}$ is the effective diameter of the solvent molecules, $y$ is the fraction of the volume occupied by the solvent molecules in the liquid ({\it i.e.}, $(1-y)$ is the fraction of the volume in between the liquid molecules), $y=(\pi/6)n_\mathrm{liq}\sigma_\mathrm{liq}^3$, $n_\mathrm{liq}$ is the solvent particle number density. In what follows we take into account that, typically, the last term in equation~\eref{eq-MCM-02} is negligible until pressure is as high as $10^8\,\mathrm{Pa}$, which is not the case for our study. Energy $G_i^{[m]}$ is yet-to-be-determined.

The condition of oil boiling at pressure $P$ is
\begin{equation}
P=\sum_m P_\mathrm{vap}^{[m]}\,.
\label{eq-MCM-03}
\end{equation}
At the boiling temperature (which depends on the oil composition), the equilibrium between the vapour and liquid phases is determined by the system of relations
\begin{equation}
Y^{[m]}P=
\frac{RT}{v_\mathrm{liq}}X^{[m]}
 \exp\left[\frac{G_c^{[m]}+G_i^{[m]}}{k_BT}\right],
\label{eq-MCM-04}
\end{equation}
where $Y^{[m]}$ is the molar fraction of specie $m$ in the vapour phase. Notice obvious relations $\sum_m X^{[m]}=1$ and $\sum_m Y^{[m]}=1$.

\subsection{Change of the mixture composition in the course of boiling}
During boiling of the mixture, the volatile components are evaporated faster than the heavy ones. Hence, the boiling temperature increases as a part of mixture is boiled-out. Within the framework of the above-described physical model, one can derive the equation system for the change of oil composition with temperature increase.

Let us assume that under pressure $P$ the boiling temperature $T_b$ of the mixture with composition $X^{[m]}$ was attained. One can calculate what number $\delta M$ of liquid molecules has to pass into the vapour phase for to raise the boiling point from $T_b$ to $T_b+\delta T$;
\begin{equation}
(M-\delta M)\overline{X}^{[m]}+\delta M\overline{Y}^{[m]}=M X^{[m]}\,,
\label{eq-MCM-05}
\end{equation}
where $M$ is the number of molecules in the liquid phase and the bar marks the new values of $X^{[m]}$ and $Y^{[m]}$. Equation~\eref{eq-MCM-04} determines relation between the changed values $\overline{X}^{[m]}$ and $\overline{Y}^{[m]}$;
\begin{equation}
\overline{Y}^{[m]}P=
\frac{R(T_b+\delta T)}{v_\mathrm{liq}+\delta v_\mathrm{liq}}\overline{X}^{[m]}
 \exp\left[\frac{G_c^{[m]}(T_b+\delta T)+G_i^{[m]}}{k_B(T_b+\delta T)}\right].
\label{eq-MCM-06}
\end{equation}
Here $v_\mathrm{liq}=\rho_\mathrm{liq}^{-1}\sum_m\mu_mX^{[m]}$, where $\mu_m$ is the effective molar mass of the specie $m$. Neglecting variation of $\rho_\mathrm{liq}$, one finds $\delta v_\mathrm{liq}=\rho_\mathrm{liq}^{-1}\sum_m\mu_m\delta X^{[m]}$.

After straightforward mathematical manipulations, one can rewrite equation~\eref{eq-MCM-05} for small variations $\delta X^{[m]}=\overline{X}^{[m]}-X^{[m]}$ and $\delta Y^{[m]}=\overline{Y}^{[m]}-Y^{[m]}$;
\begin{equation}
\delta\ln{X^{[m]}}=\left(1-\frac{Y^{[m]}}{X^{[m]}}\right)\delta\omega\,,
\label{eq-MCM-07}
\end{equation}
where $\delta\omega=\delta M/M$, which means that the current number of molecules in the liquid phase is related to the initial number of particles $M_0$ as $M=M_0\exp(-\omega)$ and the initial value of $\omega$ is $0$. Equation~\eref{eq-MCM-06} can be rewritten as
\begin{equation}
\delta Y^{[m]}=Y^{[m]}\delta\ln{X^{[m]}}
 +Y^{[m]}\left(1-\frac{G_i^{[m]}}{k_BT_b}\right)\delta\ln{T_b}
 -Y^{[m]}\frac{\sum_s\mu_s\delta X^{[s]}}{\sum_s\mu_sX^{[s]}}\,.
\label{eq-MCM-08}
\end{equation}
Substituting equation~\eref{eq-MCM-07} into \eref{eq-MCM-08} and summing over $m$, one can obtain
\[
\delta\ln{T_b}=
\frac{\displaystyle
  \sum_m\frac{\big(Y^{[m]}\big)^2}{X^{[m]}}
  -\frac{\sum_m\mu_m Y^{[m]}}{\sum_m\mu_m X^{[m]}}}
 {\displaystyle
  1+\sum_m\frac{-G_i^{[m]}}{k_BT_b}Y^{[m]}}\;\delta\omega\,.
\]
Substituting the ratio $(Y^{[m]}/X^{[m]})$ from equation~\eref{eq-MCM-04}, one finds
\begin{equation}
\delta\ln{T_b}=\frac{\displaystyle
  \sum_m\left(\frac{RT_b}{Pv_\mathrm{liq}}
  \exp\frac{G_c^{[m]}+G_i^{[m]}}{k_BT_b}\right)^2X^{[m]}
  -\frac{\displaystyle \sum_m\mu_m\frac{RT_b}{Pv_\mathrm{liq}}
         \exp\frac{G_c^{[m]}+G_i^{[m]}}{k_BT_b}X^{[m]}}
  {\displaystyle \sum_m\mu_m X^{[m]}}}
 {\displaystyle
  1\;+\;\sum_m\frac{R}{Pv_\mathrm{liq}}\frac{-G_i^{[m]}}{k_B}X^{[m]}
  \exp\frac{G_c^{[m]}+G_i^{[m]}}{k_BT_b}}
\;\delta\omega\,.
\label{eq-MCM-09}
\end{equation}
Simultaneously, equation~\eref{eq-MCM-07} reads
\begin{equation}
\delta\ln{X^{[m]}}=\left(1-\frac{RT_b}{Pv_\mathrm{liq}}
  \exp\frac{G_c^{[m]}+G_i^{[m]}}{k_BT_b}\right)\delta\omega\,.
\label{eq-MCM-10}
\end{equation}

Equations~\eref{eq-MCM-09} and \eref{eq-MCM-10} form a closed equation system for calculation of the change of $X^{[m]}$ and the relative change of the number of molecules in the liquid phase, $M/M_0=\exp(-\omega)$, as the boiling temperature increases.

\section{Evaluation of physical parameters of the system}
In this section we evaluate the physical parameters of the system from indirect empiric data.

\subsection{Saturated vapour pressure of liquid n-alkanes}
From the data on the saturated vapour pressure (including the boiling temperature) of long-chain n-alkanes, one can assess the potential energy of the interaction of the n-alkanes with the ``broth'' of the crude oil $G_i^\mathrm{[C_mH_{2m+2}]}$. The saturated vapour pressure at $293$ or $298\,\mathrm{K}$ for the substances from n-pentane to n-hexadecane and the boiling temperatures of the substances from n-hexane to n-tetradecane are well resembled by the expression
\begin{equation}
P_\mathrm{vap}^\mathrm{[C_mH_{2m+2}]}=\frac{RT}{v_\mathrm{liq}^\mathrm{[C_mH_{2m+2}]}}
 \exp\left[\gamma+\gamma_\mathrm{ln}\ln(m)-\frac{q(m)}{T}
 \right]
\label{eq-PhP-01}
\end{equation}
with $\gamma=4.178$, $\gamma_\mathrm{ln}=1.237$,
\begin{equation}
q(m)=\alpha m,\quad\alpha=500.3\,\mathrm{K};
\label{eq-PhP-02}
\end{equation}
for liquid n-alkanes $v_\mathrm{liq}^\mathrm{[C_mH_{2m+2}]}=(16.18\,m+33.00)\,\mathrm{cm^3/mol}$ at $T=273\,\mathrm{K}$ and atmospheric pressure.
Here $\gamma+\gamma_\mathrm{ln}\ln(m)$ represents entropic part of the enthalpy of molecules in the liquid phase (implies that the specific entropy is a logarithm of a power-law function of the chain length $m$), while
\[
q(m)=-G_i^\mathrm{[C_mH_{2m+2}]}/k_B\approx-G_i^{[m]}/k_B
\]
does the potential energy of the interaction of the molecule with the ambient liquid.

\subsection{Solubility of light gaseous n-alkanes in long-chain n-alkanes}
Solubility of gaseous n-alkanes in long-chain n-alkanes yields information on the effective diameter $\sigma^{[m]}$ of n-alkane molecules. On the basis of the scaled particle theory the gas solubility in liquid alkane is
\begin{equation}
X_{(0)}^\mathrm{[C_mH_{2m+2},C_lH_{2l+2}]}
 =\frac{P\,v_\mathrm{liq}^\mathrm{[C_lH_{2l+2}]}}{RT}
  \exp\left[-\frac{G_c^\mathrm{[C_mH_{2m+2},C_lH_{2l+2}]}+G_i^\mathrm{[C_mH_{2m+2},C_lH_{2l+2}]}}{k_BT}\right].
\label{eq-PhP-03}
\end{equation}
The cavity formation work~\eref{eq-MCM-02} reads
\begin{eqnarray}
\frac{G_c^\mathrm{[C_mH_{2m+2},C_lH_{2l+2}]}}{k_BT}=-\ln(1-y^\mathrm{[C_lH_{2l+2}]})
 +\frac{3y^\mathrm{[C_lH_{2l+2}]}}{1-y^\mathrm{[C_lH_{2l+2}]}}
\frac{\sigma^\mathrm{[C_mH_{2m+2}]}}{\sigma^\mathrm{[C_lH_{2l+2}]}}\qquad
\nonumber\\[5pt]
 {}+\left[\frac{3y^\mathrm{[C_lH_{2l+2}]}}{1-y^\mathrm{[C_lH_{2l+2}]}}
 +\frac{9}{2}\left(\frac{y^\mathrm{[C_lH_{2l+2}]}}{1-y^\mathrm{[C_lH_{2l+2}]}}\right)^2\right]
 \left(\frac{\sigma^\mathrm{[C_mH_{2m+2}]}}{\sigma^\mathrm{[C_lH_{2l+2}]}}\right)^2.
\label{eq-PhP-04}
\end{eqnarray}
Fitting the empiric data on the solubility of ethane and propane in long-chain alkanes~\cite{ethane-in-decane,propane-in-alkanes} at various temperatures, one finds
\begin{equation}
\sigma^{[m]}\approx\sigma^\mathrm{[C_mH_{2m+2}]}=\sigma_0 m^{1/3},
\qquad\quad\sigma_0=2.74\,\mathrm{\AA}\,.
\label{eq-PhP-05}
\end{equation}

\subsection{Solubility of gaseous n-alkanes in crude oil}
The important-for-entropy characteristics of the oil ``broth'' as a solvent, $y$ and $\sigma_\mathrm{liq}$, can be assessed from the empiric data on the solubility of light gases (methane, ethane) in crude oil.
From empiric data for crude oils~\cite{solubility-in-crude-oil}, with equations~\eref{eq-MCM-04}, \eref{eq-MCM-02}, \eref{eq-PhP-02}, \eref{eq-PhP-05}, one can find
\begin{equation}
y=0.401\,,\qquad\sigma_\mathrm{liq}=6.20\,\mathrm{\AA}\,.
\label{eq-PhP-06}
\end{equation}
With these parameters we calculate solubility determined as the number of the solute molecules per the unit volume of solvent $(X/v_\mathrm{liq})$, since this quantity (but not the molar fraction) varies insignificantly for solutions of a light gas in long-chain hydrocarbons or crude oils. For calculations with the molar fraction $X$ and equations~\eref{eq-MCM-01} or \eref{eq-MCM-04}, one has to use
\[
v_\mathrm{liq}=\frac{\sum_m\mu_mX^{[m]}}{\rho_\mathrm{liq}}\,,
\]
where $\mu_m=14m\,\mathrm{g/mol}$ is the effective molar mass of the specie $m$.

\begin{figure}[t]
\center{
\includegraphics[width=0.70\textwidth]{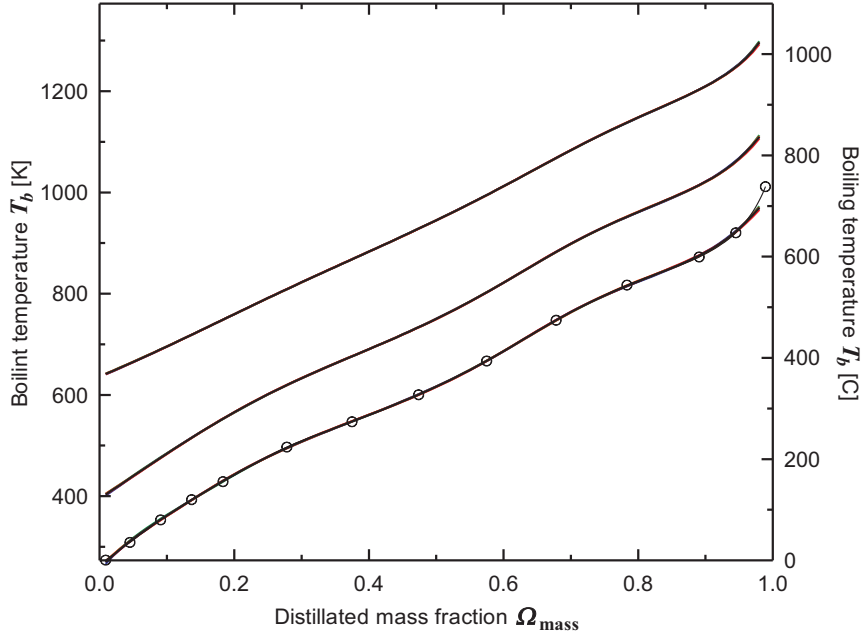}}
\caption{Temperature--distillation chart of typical crude oil: Experimental distillation chart for atmospheric pressure is plotted with circles. Bold lines represent the distillation charts calculated with the model~\eref{eq-MCM-09}--\eref{eq-MCM-10} for a reconstructed composition of crude oil at $P=1\,\mathrm{atm}$, $1\,\mathrm{MPa}$ and $10\,\mathrm{MPa}$ (from bottom to top).}
\label{fig1}
\end{figure}

\subsection{Crude oil composition and temperature--distillation chart}
Up to the authors' knowledge, the comprehensive data on the composition of crude oil (even for some sort of it), similar to data of chromatographic analysis, are not available in the literature.  However, with known parameters~\eref{eq-PhP-02}, \eref{eq-PhP-05} and \eref{eq-PhP-06}, one can use the model~\eref{eq-MCM-09}--\eref{eq-MCM-10} to reconstruct  the initial composition $X^{[m]}$ from the temperature-distillation chart---the dependence of the boiling temperature on the evaporated volume fraction of liquid oil. The volume fraction $\varOmega_\mathrm{vol}$ first needs to be recalculated into the mass fraction $\varOmega_\mathrm{mass}$;
\[
1-\varOmega_\mathrm{mass}=(1-\varOmega_\mathrm{vol})
 \frac{\rho_\mathrm{liq}(\varOmega_\mathrm{vol})}{\rho_\mathrm{liq}(\varOmega_\mathrm{vol}=0)}\,.
\]
The value $\omega$ is related to $\varOmega_\mathrm{mass}$ as follows;
\[
1-\varOmega_\mathrm{mass}=\frac{e^{-\omega}\sum_m\mu_m X^{[m]}}{\sum_m\mu_m X_\mathrm{init.}^{[m]}}\,,
\]
where $X_\mathrm{init.}^{[m]}$ is the oil composition prior to distillation.

In \fref{fig1}, one can see the results of reconstruction of the crude oil composition for a typical temperature--distillation chart. Notice, the temperature--distillation charts for elevated pressure are not results of mere rescaling or simple transformation of the chart for atmospheric pressure; the shift of the chart along temperature neither correspond to nearly logarithmic nor nearly polynomial dependence of the boiling temperature on pressure. For higher pressures the chart becomes more linear.

One should treat the results on oil composition for high values of $m$ with care. The derived composition provides correct behaviour for the early and middle stages of the distillation process; however, for the late stage, when only few components are remaining in the liquid phase, their molar fractions become non-small and one cannot rely on the theory of dilute solutions anymore. Additionally, for extremely high temperatures, chemical transformations of species become important for the distillation process. Their role increases for elevated pressure, since the boiling temperatures increases as pressure increases. Nonetheless, the subject of our primary interest is the early stages of oil distillation, where it looses light fractions. Thus, the discussed restrictions on the theory applicability are not relevant for our task even for the case of elevated pressure.

\section{Conclusion}
We have constructed the theoretical model of boiling of multicomponent mixtures of hydrocarbons (equations~\eref{eq-MCM-09}--\eref{eq-MCM-10}), calculated physical parameters of this model (equations~\eref{eq-PhP-02}, \eref{eq-PhP-05} and \eref{eq-PhP-06}) and employed them for evaluation of the crude oil composition from the temperature--distillation chart. With these data evaluated, one can numerically simulate the kinetics of crude oil boiling under elevated pressure. In particular, one can quantitatively assess the role of boiling of oil fields by lava intrusions for the light hydrocarbon gases release into the atmosphere, which is considered as a potential driving force of the Great Permean Extinction.

\ack{
Authors acknowledge financial support by the Government of Perm Region (Contract {C-26/0004.3}) and the Russian Foundation for Basic Research (project no.\ 14-01-31380\_mol\_a).}

\end{document}